\batchmode
\makeatletter
\def\input@path{{"/home/jacob/Documents/Work/My Papers/Magnetic Forces Can Do Work (2020)/"}}
\makeatother
\documentclass[twoside,twocolumn,english,prl,superscriptaddress,nobibnotes,nofootinbib]{revtex4-2}
\usepackage[latin9]{inputenc}
\usepackage{color}
\usepackage{babel}
\usepackage{mathrsfs}
\usepackage{mathtools}
\usepackage{amsmath}
\usepackage{amssymb}
\usepackage[pdftex,unicode=true,
 bookmarks=true,bookmarksnumbered=true,bookmarksopen=false,
 breaklinks=false,pdfborder={0 0 0},pdfborderstyle={},backref=false,colorlinks=true]
 {hyperref}
\hypersetup{pdftitle={On Magnetic Forces and Work},
 pdfauthor={Jacob A. Barandes},
 pdfsubject={Physics},
 pdfkeywords={Classical physics; particle physics; gauge theories; electromagnetism},
 linkcolor=blue, urlcolor=blue, citecolor=blue}

\makeatletter
\usepackage{url}

\let\originalleft\left
\let\originalright\right
\renewcommand{\left}{\mathopen{}\mathclose\bgroup\originalleft}
\renewcommand{\right}{\aftergroup\egroup\originalright}

\def\smalloverbrace#1{\mathop{\vbox{\m@th\ialign{##\crcr%
      \noalign{\kern3\p@}%
      \tiny\downbracefill\crcr\noalign{\kern3\p@\nointerlineskip}%
      $\hfil\displaystyle{#1}\hfil$\crcr}}}\limits}

\def\smallunderbrace#1{\mathop{\vtop{\m@th\ialign{##\crcr
   $\hfil\displaystyle{#1}\hfil$\crcr
   \noalign{\kern3\p@\nointerlineskip}%
   \tiny\upbracefill\crcr\noalign{\kern3\p@}}}}\limits}

\makeatother

\begin{document}
\title{On Magnetic Forces and Work}
\author{Jacob A. Barandes}
\email{jacob\_barandes@harvard.edu}

\affiliation{Jefferson Physical Laboratory, Harvard University, Cambridge, MA 02138}
\date{\today}
\begin{abstract}
We address a long-standing debate over whether classical magnetic
forces can do work, ultimately answering the question in the affirmative.
In detail, we couple a classical particle with intrinsic spin and
elementary dipole moments to the electromagnetic field, derive the
appropriate generalization of the Lorentz force law, show that the
particle's dipole moments must be collinear with its spin axis, and
argue that the magnetic field does mechanical work on the particle's
elementary magnetic dipole moment. As consistency checks, we calculate
the overall system's energy-momentum and angular momentum, and show
that their local conservation equations lead to the same force law
and therefore the same conclusions about magnetic forces and work.
We also compute the system's Belinfante-Rosenfeld energy-momentum
tensor.
\end{abstract}
\maketitle

\global\long\def\vec#1{{\bf #1}}%
\global\long\def\vecgreek#1{\boldsymbol{#1}}%
\global\long\def\dotprod{\cdot}%
\global\long\def\crossprod{\times}%
\global\long\def\tud#1#2#3{{#1^{#2}}_{#3}}%
\global\long\def\tdu#1#2#3{{#1_{#2}}^{#3}}%
\global\long\def\defeq{\equiv}%
\global\long\def\Trace{\mathrm{Tr}}%
\global\long\def\transp{\mathrm{T}}%
\global\long\def\refvalue{0}%
\global\long\def\parens#1{(#1)}%
\global\long\def\bigparens#1{\big(#1\big)}%
\global\long\def\Bigparens#1{\Big(#1\Big)}%
\global\long\def\biggparens#1{\bigg(#1\bigg)}%
\global\long\def\Biggparens#1{\Bigg(#1\Bigg)}%
\global\long\def\bracks#1{[#1]}%
\global\long\def\bigbracks#1{\big[#1\big]}%
\global\long\def\Bigbracks#1{\Big[#1\Big]}%
\global\long\def\biggbracks#1{\bigg[#1\bigg]}%
\global\long\def\Biggbracks#1{\Bigg[#1\Bigg]}%
\global\long\def\curlies#1{\{#1\}}%
\global\long\def\bigcurlies#1{\big\{#1\big\}}%
\global\long\def\Bigcurlies#1{\Big\{#1\Big\}}%
\global\long\def\biggcurlies#1{\bigg\{#1\bigg\}}%
\global\long\def\Biggcurlies#1{\Bigg\{#1\Bigg\}}%
\global\long\def\verts#1{\vert#1\vert}%
\global\long\def\bigverts#1{\big\vert#1\big\vert}%
\global\long\def\Bigverts#1{\Big\vert#1\Big\vert}%
\global\long\def\biggverts#1{\bigg\vert#1\bigg\vert}%
\global\long\def\Biggverts#1{\Bigg\vert#1\Bigg\vert}%

\section{Introduction}

Textbook treatments and research articles on classical electromagnetism,
such as \citep{Griffiths:2017ie,GrallaHarteWald:2009rdesf}, often
suggest that magnetic fields cannot do mechanical work. However, everyday
examples of bar magnets lifting other bar magnets would seem to suggest
otherwise. In this paper, we show that there exists a classical way
to understand how magnetic fields can indeed do work.\footnote{For a more extensive treatment of the results in this paper, see \citep{Barandes:2019cmfdw}.}

We start in Section~\ref{sec:The-Kinematics-of-a-Relativistic-Elementary-Dipole}
with a review of the kinematics of classical relativistic point particles
with intrinsic spin and permanent, elementary dipole moments, arguing
that these dipole moments should be collinear with the particle's
spin. In Section~\ref{sec:The-Dynamics-of-a-Relativistic-Elementary-Dipole},
we couple a particle of this kind to the electromagnetic field and
derive its dynamics, showing, in particular, that magnetic forces
can classically do work on the particle via its elementary magnetic
dipole moment, and verifying the self-consistency of the condition
that the particle's dipole moments are collinear with its spin. In
Section~\ref{sec:Conservation-Laws}, we derive expressions for the
overall system's energy-momentum and angular momentum, and show that
their associated conservation laws lead to the same equations of motion
as before, thereby providing further confirmation that magnetic fields
can do work on a particle with elementary dipole moments. We conclude
with one more new result by calculating the system's Belinfante-Rosenfeld
energy-momentum tensor.

\section{The Kinematics of a Relativistic Elementary Dipole\label{sec:The-Kinematics-of-a-Relativistic-Elementary-Dipole}}

To start, we will need a relativistic description of the kinematics
of a classical point particle with intrinsic spin.

\subsection{The Phase Space for a Relativistic Massive Particle with Spin}

The treatment of such particles has a long history\textemdash see,
for example, \citep{BargmannMichelTelegdi:1959pppmhef,HansonRegge:1974rst,BalachandranMarmoSkagerstamStern:1983gsfb,Souriau:1997sds,Rivas:2002ktsp}.
Following \citep{Barandes:2021gifcmpws,Barandes:2019mcl,SkagerstamStern:1981ldccps},
we model the particle's kinematics using spacetime coordinates $X^{\mu}=\parens{c\,T,\vec X}^{\mu}$,
relativistic-kinetic energy $E$, four-momentum $p^{\mu}=\parens{E/c,\vec p}^{\mu}$,
positive inertial mass $m>0$, and antisymmetric spin tensor $S^{\mu\nu}$
by identifying the particle's phase space as a transitive group action
(or homogeneous space) of the orthochronous Poincaré group.\footnote{This group-theoretic definition of the particle's phase space is the
classical counterpart to Wigner's classification \citep{Wigner:1939urilg}
of quantum particle-types based on irreducible Hilbert-space representations
of the Poincaré group.}

In detail, the states in this phase space take the form $\parens{X,p,S}$
and are each obtained from the unique reference state 
\begin{equation}
\parens{0,\parens{mc,\vec 0},S_{\refvalue}}\label{eq:ReferenceState}
\end{equation}
 by an appropriate Poincaré transformation $\parens{a,\Lambda}\in\mathbb{R}^{4}\ltimes O\parens{1,3}$
according to 
\begin{equation}
\parens{X,p,S}=\parens{a,\Lambda\parens{mc,\vec 0},\Lambda S_{\refvalue}\Lambda^{\transp}}.\label{eq:PhaseSpacePointFromPoincTransfFromRef}
\end{equation}
 Here the coordinates $X^{\mu}=a^{\mu}$ and the Lorentz-transformation
matrix $\tud{\Lambda}{\mu}{\nu}$, which all vary along the particle's
worldline, are treated as the particle's fundamental phase-space variables,
with the condition that $\Lambda^{\transp}\eta\Lambda=\eta=\mathrm{diag}\parens{-1,+1,+1,+1}$.
As explained in \citep{Barandes:2021gifcmpws,Barandes:2019mcl,SkagerstamStern:1981ldccps},
 the invariance of the quantities $p^{2}\defeq-m^{2}c^{2}$ and $s^{2}\defeq\parens{1/2}S_{\mu\nu}S^{\mu\nu}$
further requires the auxiliary phase-space condition 
\begin{equation}
p_{\mu}S^{\mu\nu}=0.\label{eq:FourMomSpinTensorZeroPhysicalCondition}
\end{equation}

In the reference state \eqref{eq:ReferenceState}, the particle's
four-momentum has the value $p_{\refvalue}^{\mu}=\parens{mc,\vec 0}^{\mu}=mc\,\delta_{t}^{\mu}$,
where $\delta_{\nu}^{\mu}$ is the four-dimensional Kronecker delta.
In accordance with the formula \eqref{eq:PhaseSpacePointFromPoincTransfFromRef}
for all the other states in the particle's phase space, the particle's
four-momentum in general states is therefore given in terms of the
variable Lorentz-transformation matrix $\tud{\Lambda}{\mu}{\nu}$
by $p^{\mu}=mc\,\tud{\Lambda}{\mu}t$.

Together with the reference value $p_{\refvalue}^{\mu}=\parens{mc,\vec 0}^{\mu}$
of the particle's four-momentum, the self-consistency condition \eqref{eq:FourMomSpinTensorZeroPhysicalCondition}
tells us that the value $S_{\refvalue}^{\mu\nu}$ of the particle's
spin tensor in the reference state \eqref{eq:ReferenceState} satisfies
$mc\,S_{\refvalue}^{t\nu}=0$. This equation, in turn, implies that
the reference value of the spin tensor takes the specific form 
\begin{equation}
S_{\refvalue}^{\mu\nu}=\begin{pmatrix}0 & 0 & 0 & 0\\
0 & 0 & S_{\refvalue,z} & -S_{\refvalue,y}\\
0 & -S_{\refvalue,z} & 0 & S_{\refvalue,x}\\
0 & S_{\refvalue,y} & -S_{\refvalue,x} & 0
\end{pmatrix}^{\mathclap{\mu\nu}}\label{eq:MassivePosEnergyRefSpinTensor}
\end{equation}
 for a three-dimensional pseudovector 
\begin{equation}
\vec S_{0}\defeq\parens{S_{0,x},S_{0,y},S_{0,z}}\label{eq:RefSpin3DVec}
\end{equation}
 whose direction can be chosen for convenience. As an immediate consequence,
we see that the particle's reference state \eqref{eq:ReferenceState}
spontaneously breaks the full three-dimensional rotation group down
to the symmetry subgroup of rotations around the axis defined by $\vec S_{0}$.

\subsection{Charge and Elementary Dipole Moments}

We can couple the particle to the electromagnetic field by assigning
the particle a purely electric monopole charge $q$ and an antisymmetric
elementary dipole tensor $m^{\mu\nu}$ encoding both electric and
magnetic dipole moments. The particle is then an electrically charged
elementary dipole.

Note, in particular, that the elementary magnetic dipole moments considered
in this paper are neither of the Ampère model, which would instead
consist of loops of moving electric monopoles, nor of the Gilbert
model, which would instead consist of pairs of hypothetical magnetic
monopoles or dyons. Elementary magnetic dipoles represent a classical
extension of Maxwell's original theory of electromagnetism, as Maxwell's
theory includes magnetic dipoles only of the Ampère type.\footnote{We thank David Griffiths for pointing out that a defining feature
of Maxwell's original theory is the inclusion of magnetic dipoles
solely of the Ampère type, without magnetic monopoles, dyons, Gilbert
dipoles, or elementary magnetic dipoles.} 

Note also that because elementary magnetic dipoles do not arise from
magnetic monopoles or dyons, they will not end up altering the homogeneous
Maxwell equations. Indeed, we will see that the elementary dipoles
studied in this paper correspond to derivative terms in the charge
and current densities that appear in the inhomogeneous Maxwell equations.

We let $u^{\mu}\defeq dX^{\mu}/d\lambda$ denote the particle's four-velocity
and $\gamma\defeq u^{0}/c$ denote the particle's associated Lorentz
factor, where $u^{\mu}$ is not generically normalized to $u^{2}=-c^{2}$
unless the worldline parameter $\lambda$ is taken to be the particle's
proper time $\tau$.\footnote{For maximum generality and to avoid introducing any unnecessary constraints
into the particle's Lagrangian formulation, it is convenient to wait
until after deriving the particle's equations of motion before imposing
the simplifying condition that $\lambda$ is the particle's proper
time $\tau$.} In terms of $\gamma$, $c$, and the particle's three-velocity $\vec v\defeq d\vec X/dt$,
the particle's four-velocity takes the form 
\begin{equation}
u^{\mu}=\parens{\gamma c,\gamma\vec v}^{\mu}.\label{eq:4DElementaryDipole4Vel}
\end{equation}
 The particle then has four-dimensional electric-monopole current
density 

\begin{equation}
j_{\textrm{e}}^{\nu}\parens{\vec x,t}=\parens{\rho_{\textrm{e}}\parens{\vec x,t}c,\,\vec J_{\textrm{e}}\parens{\vec x,t}}^{\nu}=qu^{\nu}\frac{1}{\gamma}\delta^{3}\parens{\vec x-\vec X}\label{eq:4DElementaryDipoleElectricCurrentDensity}
\end{equation}
 and elementary-dipole density 
\begin{equation}
M^{\mu\nu}=m^{\mu\nu}\frac{1}{\gamma}\delta^{3}\parens{\vec x-\vec X},\label{eq:ElemDipoleTensorFromDensityDeltaFunc}
\end{equation}
 which appears as a derivative contribution to the particle's total
current density, 
\begin{equation}
j^{\nu}\parens{\vec x,t}=j_{\textrm{e}}^{\nu}\parens{\vec x,t}+\partial_{\mu}M^{\mu\nu}\parens{\vec x,t},\label{eq:ElemDipoleTotalCurrentDensity}
\end{equation}
 where $\parens{1/\gamma}\delta^{3}\parens{\vec x-\vec X}$ is the
Lorentz-invariant form of the three-dimensional Dirac delta function.

It follows immediately from \eqref{eq:4DElementaryDipoleElectricCurrentDensity}
that the particle's electric-monopole density $\rho_{\textrm{e}}=j_{\textrm{e}}^{t}/c$,
its electric-monopole current density $\vec J_{\textrm{e}}=\parens{j_{\textrm{e}}^{x},j_{\textrm{e}}^{y},j_{\textrm{e}}^{z}}$,
and its three-velocity $\vec v\defeq d\vec X/dt$ satisfy the basic
relationship 
\begin{equation}
\vec J_{\textrm{e}}=\rho_{\textrm{e}}\vec v.\label{eq:ElectricMonopoleRelationChargeCurrentDensityVelocity}
\end{equation}
 We emphasize that no such relationship holds for the particle's elementary
dipole moments, which, again, are not assumed to arise as in the Ampère
model from any underlying motion of electric monopoles.

As in \citep{GrallaHarteWald:2009rdesf}, by introducing suitable
four-vectors $\pi^{\mu}$ and $\mu^{\mu}$ and antisymmetric tensors
\begin{align}
\pi^{\mu\nu} & \defeq\frac{1}{mc}\parens{p^{\mu}\pi^{\nu}-p^{\nu}\pi^{\mu}},\label{eq:4DElementaryElDipoleTensorFrom4Vecs}\\
\mu^{\mu\nu} & \defeq\frac{1}{mc}\epsilon^{\mu\nu\rho\sigma}p_{\rho}\mu_{\sigma},\label{eq:4DElementaryMagDipoleTensorFrom4Vecs}
\end{align}
 we can write the particle's elementary dipole tensor in terms of
an electric part $\pi^{\mu\nu}$ and a magnetic part $\mu^{\mu\nu}$
as 
\begin{equation}
m^{\mu\nu}=\pi^{\mu\nu}+\mu^{\mu\nu}.\label{eq:4DElementaryDipoleTensorFromElMagTensors}
\end{equation}
 Here $\epsilon^{\mu\nu\rho\sigma}$ is the four-dimensional Levi-Civita
symbol (with $\epsilon_{txyz}\defeq+1$), and the four-vectors $\pi^{\mu}$
and $\mu^{\mu}$ are related to their values in the particle's reference
state \eqref{eq:ReferenceState} and to the variable Lorentz-transformation
matrix $\tud{\Lambda}{\mu}{\nu}$ according to 
\begin{align}
\pi^{\mu} & \defeq\tud{\Lambda}{\mu}{\nu}\pi_{\refvalue}^{\nu},\label{eq:ElectricDipole4VecFromRef}\\
\mu^{\mu} & \defeq\tud{\Lambda}{\mu}{\nu}\mu_{\refvalue}^{\nu}.\label{eq:MagneticDipole4VecFromRef}
\end{align}
 We can define a three-dimensional electric-dipole vector $\vecgreek{\pi}=\parens{\pi_{x},\pi_{y},\pi_{z}}$
and a three-dimensional magnetic-dipole pseudovector $\vecgreek{\mu}=\parens{\mu_{x},\mu_{y},\mu_{z}}$
in terms of components of the elementary dipole tensor \eqref{eq:4DElementaryDipoleTensorFromElMagTensors}
according to 
\begin{equation}
m^{\mu\nu}\defeq\begin{pmatrix}0 & c\pi_{x} & c\pi_{y} & c\pi_{z}\\
-c\pi_{x} & 0 & -\mu_{z} & \mu_{y}\\
-c\pi_{y} & \mu_{z} & 0 & -\mu_{x}\\
-c\pi_{z} & -\mu_{y} & \mu_{x} & 0
\end{pmatrix}^{\mathclap{\mu\nu}}.\label{eq:4DElementaryDipoleTensorFromMoments}
\end{equation}
 Then the values $\vecgreek{\pi}_{0}$ of the electric-dipole vector
and $\vecgreek{\mu}_{0}$ of the magnetic-dipole pseudovector in the
reference state \eqref{eq:ReferenceState} are related to their four-vector
counterparts $\pi_{0}^{\mu}$ and $\mu_{0}^{\mu}$ according to 
\begin{align}
\pi_{\refvalue}^{\mu} & \defeq\parens{0,\vecgreek{\pi}_{\refvalue}}^{\mu},\label{eq:DefRefElectricDipole4VecFrom3Vec}\\
\mu_{\refvalue}^{\mu} & \defeq\parens{0,\vecgreek{\mu}_{\refvalue}}^{\mu}.\label{eq:DefRefMagneticDipole4VecFrom3Vec}
\end{align}

Recalling that the particle's reference state \eqref{eq:ReferenceState}
is unique and is symmetric under the subgroup of rotations around
the axis defined by the particle's spin pseudovector $\vec S_{0}$
from \eqref{eq:RefSpin3DVec}, we see that $\vecgreek{\pi}_{0}$ and
$\vecgreek{\mu}_{0}$ must be collinear with $\vec S_{0}$ (and therefore
must also be collinear with each other). Otherwise rotations around
$\vec S_{0}$ would alter either $\vecgreek{\pi}_{0}$ or $\vecgreek{\mu}_{0}$
(or both) and thereby lead to an infinite degeneracy incompatible
with the structure of the particle's phase space.\footnote{The quantum-mechanical analogue of this classical collinearity condition
follows from the Wigner-Eckart theorem.} 

Hence, for two constants of proportionality, $\Xi$ and $\Gamma$,
we must have the relations 
\begin{align}
\vecgreek{\pi}_{\refvalue} & =\frac{1}{c}\,\Xi\,\vec S_{\refvalue},\label{eq:RefElectricDipole3VecParallelSpin3VecCondition}\\
\vecgreek{\mu}_{\refvalue} & =\Gamma\,\vec S_{\refvalue}.\label{eq:RefMagneticDipole3VecParallelSpin3VecCondition}
\end{align}
 Because the electric dipole moment $\vecgreek{\pi}_{\refvalue}$
is a proper vector and $\vec S_{0}$ is a pseudovector, the first
of these constants, $\Xi$, must be a pseudoscalar. The magnetic dipole
moment $\vecgreek{\mu}_{0}$, by contrast, is a pseudovector, so the
other constant, $\Gamma$, must be a proper scalar. Given its relationship
to $\vecgreek{\mu}_{0}$ and $\vec S_{0}$, we can interpret $\Gamma$
as the particle's gyromagnetic ratio, whose specific value is not
fixed by our group-theoretic arguments here.

\section{The Dynamics of a Relativistic Elementary Dipole\label{sec:The-Dynamics-of-a-Relativistic-Elementary-Dipole}}

Next, we turn to a discussion of the particle's dynamics.

\subsection{The Action Functional for a Relativistic Massive Particle with Spin}

In the absence of intrinsic spin, \citep{Barandes:2021gifcmpws,Barandes:2019mcl}
show that one can always rewrite the canonical, manifestly covariant
action functional for a free relativistic particle, 
\begin{equation}
S_{\textrm{no spin}}\bracks{X,\Lambda}=\int d\lambda\,p_{\mu}\dot{X}^{\mu},\label{eq:ActionFunctionalNoSpin}
\end{equation}
 in the alternative form 
\begin{equation}
S_{\textrm{no spin}}\bracks{X,\Lambda}=\int d\lambda\,\frac{1}{2}L_{\mu\nu}\dot{\theta}^{\mu\nu},\label{eq:ActionFunctionalNoSpinFromOrbAngMom}
\end{equation}
 up to irrelevant boundary terms, where $\lambda$ is a smooth and
monotonic but otherwise arbitrary parameter along the particle's worldline,
and where 
\begin{equation}
L_{\mu\nu}\defeq X_{\mu}p_{\nu}-X_{\nu}p_{\mu}\label{eq:DefOrbitalAngMomTensor}
\end{equation}
 is the particle's orbital angular-momentum tensor.

Here $\theta^{\mu\nu}$ is an antisymmetric tensor of six independent
angular and boost variables, with $\theta^{yz},\theta^{zx},\theta^{xy}$
respectively referring to the particle's angular degrees of freedom
around the $x,y,z$ axes, and with $\theta^{tx},\theta^{ty},\theta^{tz}$
respectively referring to rapidities along the $x,y,z$ axes.\footnote{Explicitly, we have $\dot{\theta}^{\mu\nu}=\parens{i/2}\Trace\bracks{\sigma^{\mu\nu}\dot{\Lambda}\Lambda^{-1}}$,
where $\tud{\bracks{\sigma^{\mu\nu}}}{\alpha}{\beta}=-i\eta^{\mu\alpha}\delta_{\beta}^{\nu}+i\delta_{\beta}^{\mu}\eta^{\nu\alpha}$
are the generators of the Lorentz group.} These six angular and boost variables are canonically conjugate to
the corresponding six independent components of the orbital angular-momentum
tensor $L_{\mu\nu}$, with $L_{yz},L_{zx},L_{xy}$ respectively describing
the $x,y,z$ components of the particle's three-dimensional angular
momentum, and with $L_{tx},L_{ty},L_{tz}$ respectively describing
the $x,y,z$ coordinates of the particle's center of mass. 

The inclusion of intrinsic spin entails the replacement 
\begin{equation}
L_{\mu\nu}\mapsto J_{\mu\nu}=L_{\mu\nu}+S_{\mu\nu},\label{eq:TotalAngMomTensor}
\end{equation}
 where $J_{\mu\nu}$ is the particle's total angular-momentum tensor.
Continuing to assume the absence of external interactions, \citep{Barandes:2021gifcmpws,Barandes:2019mcl,SkagerstamStern:1981ldccps}
show that we can then encode the dynamics of a particle with intrinsic
spin in terms of the manifestly covariant action functional 
\begin{align}
 & S_{\textrm{particle}}\bracks{X,\Lambda}=\int d\lambda\,\frac{1}{2}J_{\mu\nu}\dot{\theta}^{\mu\nu}\nonumber \\
 & \qquad=\int d\lambda\,\biggparens{p_{\mu}\dot{X}^{\mu}+\frac{1}{2}\Trace\bracks{S\dot{\Lambda}\Lambda^{-1}}},\label{eq:ActionFunctionalWithSpin}
\end{align}
 where we again ignore irrelevant boundary terms.

\subsection{The Particle's Equations of Motion}

Our next step will be to couple the particle to the electromagnetic
field and obtain the particle's equations of motion, from which we
will be able to infer the appropriate generalization of the Lorentz
force law.

Given the charge and elementary dipole moments outlined above, the
overall action functional for the elementary dipole and the electromagnetic
field is given by 

\begin{align}
 & S\bracks{X,\Lambda,A}\defeq S_{\textrm{particle}}\bracks{X,\Lambda}+S_{\textrm{field}}\bracks A+S_{\textrm{int}}\bracks{X,\Lambda,A}\nonumber \\
\nonumber \\
 & \begin{aligned} & =\int d\lambda\,\biggparens{p_{\mu}\dot{X}^{\mu}+\frac{1}{2}\Trace\bracks{S\dot{\Lambda}\Lambda^{-1}}} & \quad & \parens{S_{\textrm{particle}}}\\
 & \qquad+\int dt\int d^{3}x\,\biggparens{-\frac{1}{4\mu_{0}}F^{\mu\nu}F_{\mu\nu}} & \quad & \parens{S_{\textrm{field}}}\\
 & \qquad+\int dt\int d^{3}x\,j^{\nu}A_{\nu} & \quad & \parens{S_{\textrm{int}}},
\end{aligned}
\label{eq:MaxwellActionWithParticle}
\end{align}
 where $F_{\mu\nu}\defeq\partial_{\mu}A_{\nu}-\partial_{\nu}A_{\mu}$
is the standard Faraday tensor and $j^{\nu}=j_{\textrm{e}}^{\nu}+\partial_{\mu}M^{\mu\nu}$
is the particle's total current density \eqref{eq:ElemDipoleTotalCurrentDensity}.
The interaction term in the final line ensures that extremizing the
action functional with respect to the electromagnetic gauge field
$A_{\mu}$ yields the Maxwell equations in their usual form, with
unmodified homogeneous equations $\nabla\dotprod\vec B=0$ and $\nabla\crossprod\vec E=-\partial\vec B/\partial t$,
and with the charge and current densities appearing in the inhomogeneous
Maxwell equations determined by \eqref{eq:ElemDipoleTotalCurrentDensity}.
The first line in this action functional ($S_{\textrm{particle}}$)
is fixed by group theory, the second line ($S_{\textrm{field}}$)
defines the vacuum in the pure Maxwell theory, and the third line
($S_{\textrm{int}}$) provides the canonical coupling between the
particle and the electromagnetic field in a manner consistent with
the Maxwell equations and the particle's features as laid out in the
previous section.

After an integration by parts, we can write the interaction term in
the final line as 
\begin{equation}
S_{\textrm{int}}\bracks{X,\Lambda,A}=\int dt\int d^{3}x\,\biggparens{j_{\textrm{e}}^{\nu}A_{\nu}-\frac{1}{2}M^{\mu\nu}F_{\mu\nu}}.\label{eq:MaxwellActionWithParticleDipoleIntsAfterIBPIntTerm}
\end{equation}
 Collecting together all the terms that involve the particle's degrees
of freedom, we obtain 
\begin{align}
 & S_{\textrm{particle+int}}\bracks{X,\Lambda,A}=\int d\lambda\,\biggparens{p_{\mu}\dot{X}^{\mu}+\frac{1}{2}\Trace\bracks{S\dot{\Lambda}\Lambda^{-1}}}\nonumber \\
 & \qquad+\int dt\int d^{3}x\,j_{\textrm{e}}^{\nu}A_{\nu}+\int dt\int d^{3}x\,\biggparens{-\frac{1}{2}}M^{\mu\nu}F_{\mu\nu},\label{eq:ParticleActionWithDipoleMomentsInitial}
\end{align}
 which we can further reduce to the form 
\begin{equation}
S_{\textrm{particle+int}}\bracks{X,\Lambda,A}=\int d\lambda\,\mathscr{L}_{\textrm{particle+int}},\label{eq:ParticleActionWithDipoleMomentsFinal}
\end{equation}
 for a manifestly covariant Lagrangian defined by 
\begin{align}
\mathscr{L}_{\textrm{particle+int}} & \defeq p_{\mu}\dot{X}^{\mu}+\frac{1}{2}\Trace\bracks{S\dot{\Lambda}\Lambda^{-1}}\nonumber \\
 & \qquad+q\dot{X}^{\nu}A_{\nu}-\frac{1}{2c}\sqrt{-\dot{X}^{2}}m^{\mu\nu}F_{\mu\nu}.\label{eq:ParticleLagrangianWithDipoleMoments}
\end{align}
 It follows from a straightforward calculation that the particle's
equations of motion, expressed in terms of the particle's proper time
$\tau$, are then 

\begin{align}
\frac{dp^{\mu}}{d\tau} & =-qu_{\nu}F^{\nu\mu}-\frac{1}{2}m^{\rho\sigma}\partial^{\mu}F_{\rho\sigma}-\frac{1}{2c^{2}}\frac{d}{d\tau}\parens{u^{\mu}m^{\rho\sigma}F_{\rho\sigma}}\nonumber \\
 & =-qu_{\nu}F^{\nu\mu}-\frac{1}{2}m^{\rho\sigma}\parens{\eta^{\mu\nu}+\frac{1}{c^{2}}u^{\mu}u^{\nu}}\partial_{\nu}F_{\rho\sigma}\nonumber \\
 & \qquad\qquad-\frac{1}{2c^{2}}\frac{d}{d\tau}\parens{u^{\mu}m^{\rho\sigma}}F_{\rho\sigma},\label{eq:ParticleWithDipoleMomentsCoordEOMProperTime}
\end{align}
 as obtained in \citep{VanDamRuijgrok:1980crepsmef,SkagerstamStern:1981ldccps,GerochWeatherall:2018msbst},
and 
\begin{equation}
\frac{dS^{\mu\nu}}{d\tau}=-\parens{u^{\mu}p^{\nu}-u^{\nu}p^{\mu}}-\parens{m^{\mu\rho}\tud F{\nu}{\rho}-m^{\nu\rho}\tud F{\mu}{\rho}},\label{eq:ParticleSpinTensorEOMProperTime}
\end{equation}
 which generalizes the results of \citep{BargmannMichelTelegdi:1959pppmhef,VanDamRuijgrok:1980crepsmef,SkagerstamStern:1981ldccps}.

\subsection{The Non-Relativistic Limit with Time-Independent External Fields}

In the non-relativistic limit for time-independent fields, and ignoring
self-field effects\textemdash so that we can regard the overall electric
and magnetic fields as external fields $\vec E_{\textrm{ext}}$ and
$\vec B_{\textrm{ext}}$\textemdash the equations of motion \eqref{eq:ParticleWithDipoleMomentsCoordEOMProperTime}
and \eqref{eq:ParticleSpinTensorEOMProperTime} reduce to 
\begin{align}
\frac{dE}{dt} & \approx\frac{d}{dt}\parens{-q\Phi_{\textrm{ext}}+\vecgreek{\pi}\dotprod\vec E_{\textrm{ext}}+\vecgreek{\mu}\dotprod\vec B_{\textrm{ext}}},\label{eq:NonrelEnergyLawOnDipole}\\
\frac{d\vec p}{dt} & \approx q\parens{\vec E_{\textrm{ext}}+\vec v\crossprod\vec B_{\textrm{ext}}}+\nabla\parens{\vecgreek{\pi}\dotprod\vec E_{\textrm{ext}}+\vecgreek{\mu}\dotprod\vec B_{\textrm{ext}}},\label{eq:NonrelForceLawOnDipole}\\
\frac{d\vec J}{dt} & \approx\vec X\crossprod\frac{d\vec p}{dt}+\vec{\vecgreek{\pi}}\crossprod\vec E_{\textrm{ext}}+\vec{\vecgreek{\mu}}\crossprod\vec B_{\textrm{ext}}.\label{eq:NonrelTorqueEqOnDipole}
\end{align}
 Here the electric field $\vec E_{\textrm{ext}}$ is given in terms
of the scalar potential $\Phi_{\textrm{ext}}$ according to the usual
formula $\vec E_{\textrm{ext}}=-\nabla\Phi_{\textrm{ext}}$ appropriate
to the static case, the particle's four-momentum in this limit is
\begin{equation}
p^{\mu}=\parens{E/c,\vec p}^{\mu}\approx\parens{mc+\parens{1/2}m\vec v^{2}/c,\vec p}^{\mu},\label{eq:NonrelParticleFourMom}
\end{equation}
 and the particle's total angular-momentum pseudovector $\vec J$
is made up of orbital and spin contributions according to 
\begin{equation}
\vec J\defeq\vec L+\vec S=\parens{L^{yz},L^{zx},L^{xy}}+\parens{S^{yz},S^{zx},S^{xy}}.\label{eq:ParticleTotAngMom3VecFromOrbSpin}
\end{equation}

 The dynamical equation \eqref{eq:NonrelEnergyLawOnDipole} for the
rate of change of the particle's relativistic-kinetic energy $E$
describes conservation of the particle's total energy, provided that
we identify the combination 
\begin{equation}
V=q\Phi_{\textrm{ext}}-\vecgreek{\pi}\dotprod\vec E_{\textrm{ext}}-\vecgreek{\mu}\dotprod\vec B_{\textrm{ext}}\label{eq:NonrelPotentialEnergy}
\end{equation}
 as the particle's potential energy. Meanwhile, the dynamical equation
\eqref{eq:NonrelForceLawOnDipole} tells us that the electromagnetic
force on the particle is 
\begin{equation}
\vec F=q\vec E_{\textrm{ext}}+q\vec v\crossprod\vec B_{\textrm{ext}}+\nabla\parens{\vecgreek{\pi}\dotprod\vec E_{\textrm{ext}}}+\nabla\parens{\vecgreek{\mu}\dotprod\vec B_{\textrm{ext}}}.\label{eq:LorentzForceLawWithDipoles}
\end{equation}
 We observe that the usual Lorentz force law, $q\vec E_{\textrm{ext}}+q\vec v\crossprod\vec B_{\textrm{ext}}$,
is enhanced in the presence of the particle's elementary dipole moments
by the appearance of two additional dipole terms $\nabla\parens{\vecgreek{\pi}\dotprod\vec E_{\textrm{ext}}}$
and $\nabla\parens{\vecgreek{\mu}\dotprod\vec B_{\textrm{ext}}}$.
The magnetic field therefore contributes to the work done by the external
electromagnetic field, $W\defeq\int d\vec X\dotprod\vec F$: 
\begin{equation}
W=\int_{A}^{B}dt\,\parens{q\vec v\dotprod\vec E_{\textrm{ext}}}+\Delta\parens{\vecgreek{\pi}\dotprod\vec E_{\textrm{ext}}}+\Delta\parens{\vecgreek{\mu}\dotprod\vec B_{\textrm{ext}}}.\label{eq:EMWorkDoneOnParticle}
\end{equation}

We have reached the key conclusion of this paper\textemdash namely,
that magnetic forces can do work on classical particles with elementary
dipole moments.\footnote{We thank Sebastiano Covone for suggesting the relevance of these results
to the Bohr-van Leeuwen theorem \citep{BohrRosenfeldNielsen:1972ddtt,VanLeeuwen:1921crepsmef}.
The Bohr-van Leeuwen theorem assumes the original Lorentz force law
without contributions from elementary dipole moments, and asserts
that a non-rotating system of particles, when treated classically,
always has a vanishing average magnetization at thermal equilibrium.
The theorem's implication is that phenomena like diamagnetism can
only be understood in terms of quantum effects, a view challenged
by our results, at least in principle.} We next turn to a detailed treatment of self-consistency conditions
on the particle's dynamics, as well as obtain the necessary formulas
for determining the particle's four-velocity $u^{\mu}$ in the presence
of a nonzero electromagnetic field. Later on, we will analyze electromagnetic
forces and work done on the particle from the standpoint of local
conservation laws.

\subsection{Implications of Self-Consistency}

Taking a derivative of the phase-space condition $p_{\mu}S^{\mu\nu}$
from \eqref{eq:FourMomSpinTensorZeroPhysicalCondition} yields the
self-consistency requirement 

\[
\frac{dp_{\mu}}{d\tau}S^{\mu\nu}+p_{\mu}\frac{dS^{\mu\nu}}{d\tau}=0.
\]
 Together with \eqref{eq:ParticleSpinTensorEOMProperTime}, this self-consistency
requirement entails that the particle's four-momentum $p^{\mu}$ and
its four-velocity $u^{\mu}=dX^{\mu}/d\tau$ (now normalized to $u^{2}=-c^{2}$)
are related by 
\begin{equation}
p^{\mu}=m_{\textrm{eff}}u^{\mu}+b^{\mu}.\label{eq:FourMomFrom4VelPlusDiscrep}
\end{equation}
 Here $m_{\textrm{eff}}$, which plays the role of an effective mass,
is defined by 
\begin{equation}
m_{\textrm{eff}}\defeq-\frac{m^{2}c^{2}}{p\dotprod u},\label{eq:DefEffectiveMass}
\end{equation}
 and the four-vector $b^{\mu}$, which is orthogonal to the particle's
four-momentum, $b\dotprod p=0$, is given by 
\begin{equation}
b^{\mu}\defeq\frac{1}{p\dotprod u}\biggparens{\frac{dp_{\nu}}{d\tau}S^{\nu\mu}-p_{\nu}\parens{m^{\nu\rho}\tud F{\mu}{\rho}-m^{\mu\rho}\tud F{\nu}{\rho}}}.\label{eq:Def4VecDiscrep}
\end{equation}
As in \citep{SkagerstamStern:1981ldccps}, we regard \eqref{eq:FourMomFrom4VelPlusDiscrep}
as an implicit\emph{ }formula for determining the behavior of the
particle's four-velocity $u^{\mu}$ as a function of the proper time.
This formula ensures, in particular, that the particle's four-momentum
$p^{\mu}$ has constant norm-squared $p^{2}=-m^{2}c^{2}$.

For vanishing field, $F_{\mu\nu}=0$, the relationship \eqref{eq:FourMomFrom4VelPlusDiscrep}
reduces to the familiar equation $p^{\mu}=mu^{\mu}$, as expected.
By contrast, when the electromagnetic field is nonzero, $F_{\mu\nu}\ne0$,
\eqref{eq:FourMomFrom4VelPlusDiscrep} has the form 
\begin{equation}
p^{\mu}=mu^{\mu}+\parens{\textrm{terms of order \ensuremath{1/c^{2}}}}.\label{eq:FourMomEqFourVelUpToRelCorrections}
\end{equation}
 This relation ensures that there is no ambiguity over whether we
should identify the particle's relativistic-kinetic energy $E$ as
$p^{t}c$ or $u^{t}mc^{2}$ for the purposes of quantifying the work
done by the field on the particle in the non-relativistic regime.

Invoking the spin tensor's equation of motion \eqref{eq:ParticleSpinTensorEOMProperTime},
together with the phase-space condition \eqref{eq:FourMomSpinTensorZeroPhysicalCondition},
$p_{\mu}S^{\mu\nu}=0$, and the constancy of the particle's spin-squared
scalar $s^{2}\defeq\parens{1/2}S_{\mu\nu}S^{\mu\nu}$, we find 
\begin{align}
\frac{d}{d\tau}\parens{s^{2}} & =\frac{d}{d\tau}\biggparens{\frac{1}{2}S_{\mu\nu}S^{\mu\nu}}\nonumber \\
 & =\parens{\tud S{\rho}{\mu}m^{\mu\sigma}-\tud S{\sigma}{\mu}m^{\mu\rho}}F_{\rho\sigma}=0,\label{eq:SpinTensorSquareConst}
\end{align}
 which yields the condition 
\begin{equation}
\tud S{\rho}{\mu}m^{\mu\sigma}=\tud S{\sigma}{\mu}m^{\mu\rho}.\label{eq:SpinTensorDipoleTensorCondition}
\end{equation}
 In the particle's reference state \eqref{eq:ReferenceState}, this
equality produces the relations 
\begin{align}
\vecgreek{\pi}_{\refvalue}\crossprod\vec S_{\refvalue} & =0,\label{eq:RefElectricDipole3VecCrossProdSpin3VecZeroCondition}\\
\vecgreek{\mu}_{\refvalue}\crossprod\vec S_{\refvalue} & =0,\label{eq:RefMagneticDipole3VecCrossProdSpin3VecZeroCondition}
\end{align}
 which ensure self-consistency with our requirement that the particle's
elementary electric and magnetic dipole moments must be collinear
with the particle's spin pseudovector $\vec S_{\refvalue}$. Physically
speaking, we can understand the self-consistency conditions \eqref{eq:RefElectricDipole3VecCrossProdSpin3VecZeroCondition}
as telling us that if the particle's elementary-dipole vectors were
not collinear with the particle's spin axis, then torques exerted
on the particle by the electromagnetic field would cause the particle's
spin to speed up or slow down, in violation of the constancy of $s^{2}$.

\section{Conservation Laws\label{sec:Conservation-Laws}}

For completeness, we verify that the equations of motion \eqref{eq:ParticleWithDipoleMomentsCoordEOMProperTime}
and \eqref{eq:ParticleSpinTensorEOMProperTime} also follow from local
conservation of energy-momentum and angular momentum. To begin, we
recall the relevant version of Noether's theorem (see, for instance,
\citep{Barandes:2019cmfdw}), which states that if a system's dynamics
has a continuous symmetry, 
\begin{align}
q_{\alpha} & \mapsto q_{\alpha}^{\prime}=q_{\alpha}+\delta_{\epsilon}q_{\alpha},\nonumber \\
 & \qquad\qquad\delta_{\epsilon}q_{\alpha}=\sum_{b}g_{q_{\alpha},b}\epsilon_{b},\label{eq:InfinitesimalSymmetryTransf}
\end{align}
 where the quantities $\epsilon_{b}$ parameterize the symmetry and
the quantities $g_{q_{\alpha},b}$ characterize its precise form,
then we have the following conservation law: 
\begin{equation}
Q_{b}\defeq\sum_{\alpha}\frac{\partial L}{\partial\dot{q}_{\alpha}}g_{q_{\alpha},b}-f_{b},\quad\frac{dQ_{b}}{dt}=0.\label{eq:NoethersTheorem}
\end{equation}
 Here $Q_{b}$ are a set of conserved quantities, $L$ is the system's
Lagrangian, $q_{\alpha}$ are its degrees of freedom, and the functions
$f_{b}$ are related to the change in the Lagrangian according to
\begin{align}
L & \mapsto L+\delta_{\epsilon}L,\nonumber \\
 & \qquad\delta_{\epsilon}L=\frac{d}{dt}\biggparens{\sum_{b}f_{b}\epsilon_{b}}=\sum_{b}\frac{df_{b}}{dt}\epsilon_{b}.\label{eq:InfinitesimalSymmetryChangeLagrangian}
\end{align}

\subsection{Local Conservation of Energy-Momentum}

In order to employ Noether's theorem to obtain the overall system's
energy-momentum tensor, we examine the behavior of the system under
a translation in spacetime by an infinitesimal four-vector $\epsilon^{\mu}$.
The particle's phase-space variables transform as 

\begin{equation}
\left.\begin{aligned}X^{\mu}\parens{\lambda} & \mapsto X^{\prime\mu}\parens{\lambda}\defeq X^{\mu}\parens{\lambda}+\epsilon^{\mu},\\
\tud{\Lambda}{\mu}{\nu}\parens{\lambda} & \mapsto\tud{\Lambda}{\prime\mu}{\nu}\parens{\lambda}\defeq\tud{\Lambda}{\mu}{\nu}\parens{\lambda},
\end{aligned}
\quad\right\} \label{eq:SpacetimeTranslationForNoetherParticleDOF}
\end{equation}
 and the electromagnetic gauge potential transforms as 
\begin{align}
A_{\mu}\parens x & \mapsto A_{\mu}^{\prime}\parens x\defeq A_{\mu}\parens{x-\epsilon}\nonumber \\
 & \qquad=A_{\mu}\parens x-\partial_{\nu}A_{\mu}\parens x\epsilon^{\nu}.\label{eq:SpacetimeTranslationForNoetherGaugeField}
\end{align}

By an application of Noether's theorem to the particle's manifestly
covariant Lagrangian $\mathscr{L}\defeq\mathscr{L}_{\textrm{particle+int}}$
defined by \eqref{eq:ParticleActionWithDipoleMomentsFinal} and the
Lagrangian density $\mathcal{L}$ for the overall system defined in
terms of the action functional $S\bracks{X,\Lambda,A}\defeq\int dt\int d^{3}x\,\mathcal{L}$
from \eqref{eq:MaxwellActionWithParticle}, one finds that the overall
system's conserved four-momentum is expressible as 
\begin{align}
P_{\nu} & =\frac{\partial\mathscr{L}}{\partial\dot{X}^{\rho}}g_{X^{\rho},\nu}+\int d^{3}x\,\parens{-n_{\mu}}\frac{\partial\mathcal{L}}{\partial\parens{c\partial_{\mu}A_{\rho}}}g_{A_{\rho},\nu}-f_{\nu}\nonumber \\
 & =p_{\nu}+qA_{\nu}+\frac{1}{2c^{2}}u_{\nu}m^{\sigma\tau}F_{\sigma\tau}\nonumber \\
 & \quad+\frac{1}{c}\int d^{3}x\,\parens{-n_{\mu}}\biggparens{H^{\mu\rho}\partial_{\nu}A_{\rho}-\delta_{\nu}^{\mu}\biggparens{\frac{1}{4\mu_{0}}F^{\rho\sigma}F_{\rho\sigma}}}\nonumber \\
 & =\frac{1}{c}\int d^{3}x\,\parens{-n_{\mu}}T_{\textrm{can},\nu}^{\mu},\label{eq:TotalCanonicalMomentumFromEnergyMomTensor}
\end{align}
 where $n_{\mu}\defeq\parens{-1,\vec 0}_{\mu}$ is a unit timelike
four-vector orthogonal to the three-dimensional spatial hypersurface
of integration. In this expression, the overall system's canonical
energy-momentum tensor is given by 
\begin{equation}
T_{\textrm{can}}^{\mu\nu}=T_{\textrm{can},\textrm{particle}}^{\mu\nu}+T_{\textrm{can},\textrm{field}}^{\mu\nu},\label{eq:TotalCanonicalEnergyMomentumTensor}
\end{equation}
 with the contributions from the particle and the field given respectively
by\footnote{The authors of \citep{GrallaHarteWald:2009rdesf} decompose the overall
energy-momentum tensor by including the interaction terms with the
energy-momentum tensor for the \emph{particle}, an approach that obscures
the work being done by the electromagnetic field on the particle.} 
\begin{equation}
T_{\textrm{can},\textrm{particle}}^{\mu\nu}\defeq u^{\mu}p^{\nu}\frac{1}{\gamma}\delta^{3}\parens{\vec x-\vec X}\label{eq:CanonicalEnergyMomentumTensorParticle}
\end{equation}
 and 
\begin{align}
T_{\textrm{can},\textrm{field}}^{\mu\nu} & \defeq H^{\mu\rho}\tud F{\nu}{\rho}-\eta^{\mu\nu}\frac{1}{4\mu_{0}}F^{2}\nonumber \\
 & \qquad+\frac{1}{2c^{2}}u^{\mu}u^{\nu}m^{\rho\sigma}F_{\rho\sigma}\frac{1}{\gamma}\delta^{3}\parens{\vec x-\vec X}\nonumber \\
 & \qquad+\partial_{\rho}\parens{H^{\mu\rho}A^{\nu}}.\label{eq:CanonicalEnergyMomentumTensorEMField}
\end{align}
 Here $H^{\mu\nu}$ is the auxiliary Faraday tensor: 
\begin{align}
H^{\mu\nu} & \defeq\frac{1}{\mu_{0}}F^{\mu\nu}+M^{\mu\nu}\nonumber \\
 & =\frac{1}{\mu_{0}}F^{\mu\nu}+m^{\mu\nu}\frac{1}{\gamma}\delta^{3}\parens{\vec x-\vec X}.\label{eq:4DAuxiliaryFaradayTensorForDipoleParticle}
\end{align}
 The last term in \eqref{eq:CanonicalEnergyMomentumTensorEMField}
is a total spacetime divergence with vanishing divergence $\partial_{\mu}\partial_{\rho}\parens{H^{\mu\rho}A^{\nu}}=0$
on its $\mu$ index, and its temporal component $\partial_{\rho}\parens{H^{t\rho}A^{\nu}}$
has vanishing integral over three-dimensional space under the assumption
that the fields go to zero sufficiently rapidly at spatial infinity.
We emphasize that in our approach, all the terms in the overall system's
canonical energy-momentum tensor follow from the systematic application
of Noether's theorem to the relevant action functionals.

We can integrate the local conservation law $\partial_{\mu}T_{\textrm{can}}^{\mu\nu}=0$
over three-dimensional space to compute the time derivative of the
particle's four-momentum $p^{\nu}$: 
\begin{align*}
\frac{dp^{\nu}}{dt} & =\frac{1}{c}\frac{d}{dt}\int d^{3}x\,T_{\textrm{can},\textrm{particle}}^{t\nu}\\
 & =-\frac{1}{c}\frac{d}{dt}\int d^{3}x\,T_{\textrm{can},\textrm{field}}^{t\nu}\\
 & =\int d^{3}x\,\biggparens{-\partial_{\mu}\biggparens{H^{\mu\rho}\tud F{\nu}{\rho}-\eta^{\mu\nu}\frac{1}{4\mu_{0}}F^{2}}}\\
 & \qquad-\frac{1}{2c^{2}}\frac{d}{dt}\parens{u^{\nu}m^{\rho\sigma}F_{\rho\sigma}}\\
 & =-qu_{\mu}F^{\mu\nu}+m_{\rho\mu}\partial^{\mu}F^{\nu\rho}-\frac{1}{2c^{2}}\frac{d}{d\tau}\parens{u^{\nu}m^{\rho\sigma}F_{\rho\sigma}}.
\end{align*}
 After invoking the electromagnetic Bianchi identity $\partial^{\mu}F^{\nu\rho}+\partial^{\rho}F^{\mu\nu}+\partial^{\nu}F^{\rho\mu}=0$,
we obtain the equation of motion \eqref{eq:ParticleWithDipoleMomentsCoordEOMProperTime}. 

Our formulas above for the overall system's canonical energy-momentum
tensor are new results. By replicating the particle's equation of
motion \eqref{eq:ParticleWithDipoleMomentsCoordEOMProperTime}, they
provide further support for the key claim of this paper\textemdash that
magnetic forces can classically do work on particles with elementary
dipole moments.

\subsection{Local Conservation of Angular Momentum}

Next, we use Noether's theorem to examine the overall system's angular
momentum and its local conservation. Under an infinitesimal Lorentz
transformation 

\begin{equation}
\Lambda_{\textrm{inf}}=1+\frac{i}{2}\epsilon^{\rho\sigma}\sigma_{\rho\sigma},\label{eq:InfinitesimalLorentzTransfFromGeneratorsForNoetherElectromagnetism}
\end{equation}
 the particle's phase-space variables transform as 
\begin{equation}
\left.\begin{aligned}X^{\mu}\parens{\lambda} & \mapsto X^{\prime\mu}\parens{\lambda}\defeq\parens{\Lambda_{\textrm{inf}}X\parens{\lambda}}^{\mu}\\
 & \qquad=X^{\mu}\parens{\lambda}+\frac{i}{2}\epsilon^{\rho\sigma}\tud{\bracks{\sigma_{\rho\sigma}}}{\mu}{\nu}X^{\nu}\parens{\lambda},\\
\tud{\Lambda}{\mu}{\nu}\parens{\lambda} & \mapsto\tud{\Lambda}{\prime\mu}{\nu}\parens{\lambda}\defeq\tud{\parens{\Lambda_{\textrm{inf}}\Lambda\parens{\lambda}}}{\mu}{\nu}\\
 & \qquad=\tud{\Lambda}{\mu}{\nu}\parens{\lambda}+\frac{i}{2}\epsilon^{\rho\sigma}\tud{\bracks{\sigma_{\rho\sigma}}}{\mu}{\lambda}\tud{\Lambda}{\lambda}{\nu}\parens{\lambda}.
\end{aligned}
\quad\right\} \label{eq:LorentzTranslationForNoetherParticleDOF}
\end{equation}
 The second of these two transformation laws is equivalent to the
following transformation rule for the particle's Lorentz parameters
$\theta^{\mu\nu}\parens{\lambda}$: 
\begin{equation}
\theta^{\mu\nu}\parens{\lambda}\mapsto\theta^{\prime\mu\nu}\parens{\lambda}\defeq\theta^{\mu\nu}\parens{\lambda}+\epsilon^{\mu\nu}.\label{eq:LorentzTranslationForNoetherParticleBoostAngleDOF}
\end{equation}
 Meanwhile, the gauge field $A_{\mu}\parens x$ transforms as 
\begin{align}
A_{\mu}\parens x & \mapsto A_{\mu}^{\prime}\parens x\defeq\parens{A\parens{\Lambda_{\textrm{inf}}^{-1}x}\Lambda_{\textrm{inf}}^{-1}}_{\mu}\nonumber \\
 & \defeq A_{\lambda}\parens{\parens{1-\parens{i/2}\epsilon^{\rho\sigma}\sigma_{\rho\sigma}}x}\parens{\delta_{\mu}^{\lambda}-\parens{i/2}\epsilon^{\rho\sigma}\tud{\bracks{\sigma_{\rho\sigma}}}{\lambda}{\mu}}\nonumber \\
 & =A_{\mu}\parens x-\partial_{\nu}A_{\mu}\parens x\parens{i/2}\epsilon^{\rho\sigma}\tud{\bracks{\sigma_{\rho\sigma}}}{\nu}{\lambda}x^{\lambda}\nonumber \\
 & \qquad-A_{\lambda}\parens x\parens{i/2}\epsilon^{\rho\sigma}\tud{\bracks{\sigma_{\rho\sigma}}}{\lambda}{\mu}.\label{eq:LorentzTranslationForNoetherGaugeField}
\end{align}

Noether's theorem \eqref{eq:NoethersTheorem} then yields the system's
overall angular-momentum tensor, up to an overall minus sign: 
\begin{align}
 & -J_{\nu\rho}=\frac{\partial\mathscr{L}}{\partial\dot{X}^{\alpha}}g_{X^{\alpha},\nu\rho}+\frac{1}{2}\frac{\partial\mathscr{L}}{\partial\dot{\theta}^{\alpha\beta}}g_{\theta^{\alpha\beta},\nu\rho}\nonumber \\
 & \qquad+\int d^{3}x\,\parens{-n_{\mu}}\frac{\partial\mathcal{L}}{\partial\parens{c\partial_{\mu}A_{\alpha}}}g_{A_{\alpha},\nu\rho}-f_{\nu\rho}\nonumber \\
 & =-\biggparens{p_{\alpha}+qA_{\alpha}-\frac{1}{2}\parens{-u_{\alpha}/c^{2}}m^{\sigma\lambda}F_{\sigma\lambda}}\parens{X_{\nu}\delta_{\rho}^{\alpha}-X_{\rho}\delta_{\nu}^{\alpha}}\nonumber \\
 & \quad-S_{\nu\rho}\nonumber \\
 & \quad-\frac{1}{c}\int d^{3}x\,\parens{-n_{\mu}}\biggparens{H^{\mu\alpha}-\delta_{\sigma}^{\mu}\biggparens{\frac{1}{4\mu_{0}}F^{2}}}\nonumber \\
 & \qquad\qquad\qquad\times\partial_{\sigma}A_{\alpha}\parens{x_{\nu}\delta_{\rho}^{\sigma}-x_{\rho}\delta_{\nu}^{\sigma}}\nonumber \\
 & \quad-\frac{1}{c}\int d^{3}x\,\parens{-n_{\mu}}\parens{\tud H{\mu}{\nu}A_{\rho}-\tud H{\mu}{\rho}A_{\nu}}\nonumber \\
 & =-\int d^{3}x\,\parens{-n_{\mu}}\mathcal{J}_{\textrm{can},\nu\rho}^{\mu}.\label{eq:TotalAngMomentumFromAngMomTensor}
\end{align}
 Here we have identified the system's canonical angular-momentum flux
tensor as 
\begin{equation}
\mathcal{J}_{\textrm{can}}^{\mu\nu\rho}=\mathcal{L}^{\mu\nu\rho}+\mathcal{S}^{\mu\nu\rho},\label{eq:TotalCanonicalAngMomFluxTensorExplicit}
\end{equation}
 with orbital contribution 
\begin{equation}
\mathcal{L}^{\mu\nu\rho}\defeq x^{\nu}\frac{1}{c}T_{\textrm{can}}^{\mu\rho}-x^{\rho}\frac{1}{c}T_{\textrm{can}}^{\mu\nu}\label{eq:OrbAngFluxTensorEMAndParticle}
\end{equation}
 and spin contribution 
\begin{equation}
\mathcal{S}^{\mu\nu\rho}=\frac{1}{c}u^{\mu}S^{\nu\rho}\frac{1}{\gamma}\delta^{3}\parens{\vec x-\vec X}+\frac{1}{c}\parens{H^{\mu\nu}A^{\rho}-H^{\mu\rho}A^{\nu}}.\label{eq:SpinFluxTensorEMAndParticle}
\end{equation}
 We naturally read off the spin flux tensors for the particle and
the field respectively as 
\begin{align}
\mathcal{S}_{\textrm{particle}}^{\mu\nu\rho} & =\frac{1}{c}u^{\mu}S^{\nu\rho}\frac{1}{\gamma}\delta^{3}\parens{\vec x-\vec X},\label{eq:ParticleSpinAngMomTensor}\\
\mathcal{S}_{\textrm{field}}^{\mu\nu\rho} & =\frac{1}{c}\parens{H^{\mu\nu}A^{\rho}-H^{\mu\rho}A^{\nu}}.\label{eq:SpinFluxTensorEMAlone}
\end{align}
 Integrating the local conservation law $\partial_{\mu}\mathcal{J}_{\textrm{can}}^{\mu\nu\rho}=0$
over three-dimensional space and taking advantage of the local conservation
$\partial_{\mu}T_{\textrm{can}}^{\mu\rho}=0$ of the overall canonical
energy-momentum tensor $T_{\textrm{can}}^{\mu\rho}$, we can compute
the time derivative of the particle's spin tensor as follows: 
\begin{align*}
 & \frac{dS^{\nu\rho}}{dt}=\frac{d}{dt}\int d^{3}x\,\mathcal{S}_{\textrm{particle}}^{t\nu\rho}\\
 & \quad=-\frac{d}{dt}\int d^{3}x\,\frac{1}{c}\parens{x^{\nu}T_{\textrm{can}}^{t\rho}-x^{\rho}T_{\textrm{can}}^{t\nu}+H^{t\nu}A^{\rho}-H^{t\rho}A^{\nu}}\\
 & \quad=-\int d^{3}x\,\partial_{\mu}\parens{x^{\nu}T_{\textrm{can}}^{\mu\rho}-x^{\rho}T_{\textrm{can}}^{\mu\nu}+H^{\mu\nu}A^{\rho}-H^{\mu\rho}A^{\nu}}\\
 & \quad=-\frac{1}{\gamma}\parens{u^{\nu}p^{\rho}-u^{\rho}p^{\nu}}-\frac{1}{\gamma}\parens{m^{\nu\sigma}\tud F{\rho}{\sigma}-m^{\rho\sigma}\tud F{\nu}{\sigma}}.
\end{align*}
 We therefore see that local conservation of angular momentum yields
the equation of motion \eqref{eq:ParticleSpinTensorEOMProperTime}.

\subsection{The Belinfante-Rosenfeld Energy-Momentum Tensor}

The overall system's canonical energy-momentum tensor \eqref{eq:TotalCanonicalEnergyMomentumTensor}
is not symmetric on its two indices, a feature that is required of
the energy-momentum tensor that locally sources the gravitational
field in general relativity. To conclude this paper, we follow the
standard Belinfante-Rosenfeld construction\footnote{For a review, see \citep{DiFrancescoMathieuSenechal:1996cft}.}
to construct a properly symmetric energy-momentum tensor, which will
likewise represent a new result.

We start by introducing a new tensor 

\begin{align}
 & \mathcal{B}^{\mu\rho\nu}\defeq\frac{c}{2}\parens{\mathcal{S}^{\mu\nu\rho}+\mathcal{S}^{\nu\mu\rho}+\mathcal{S}^{\rho\mu\nu}}\nonumber \\
 & =-H^{\mu\rho}A^{\nu}+\frac{1}{2}\parens{u^{\mu}S^{\nu\rho}+u^{\nu}S^{\mu\rho}+u^{\rho}S^{\mu\nu}}\frac{1}{\gamma}\delta^{3}\parens{\vec x-\vec X}.\label{eq:BelinfanteRosenfeldCorrectionEMAndParticle}
\end{align}
 We then obtain a symmetric, locally conserved energy-momentum tensor
$T^{\mu\nu}$ for the overall system from the relation $T^{\mu\nu}=T_{\textrm{can}}^{\mu\nu}+\partial_{\rho}\mathcal{B}^{\mu\rho\nu}$:\footnote{This formula differs from the corresponding result in \citep{VanDamRuijgrok:1980crepsmef},
whose energy-momentum tensor yields the correct equations of motion
for the particle only after an unjustified four-dimensional integration
by parts.} 
\begin{align}
T^{\mu\nu} & =\frac{1}{2}\parens{u^{\mu}p^{\nu}+u^{\nu}p^{\mu}}\frac{1}{\gamma}\delta^{3}\parens{\vec x-\vec X}\nonumber \\
 & \qquad+\frac{1}{2}H^{\mu\rho}\tud F{\nu}{\rho}+\frac{1}{2}H^{\nu\rho}\tud F{\mu}{\rho}-\eta^{\mu\nu}\frac{1}{4\mu_{0}}F^{\rho\sigma}F_{\rho\sigma}\nonumber \\
 & \qquad+\frac{1}{2c^{2}}u^{\mu}u^{\nu}m^{\rho\sigma}F_{\rho\sigma}\frac{1}{\gamma}\delta^{3}\parens{\vec x-\vec X}\nonumber \\
 & \qquad+\frac{1}{2}\partial_{\rho}\parens{\mathcal{S}_{\textrm{particle}}^{\mu\nu\rho}+\mathcal{S}_{\textrm{particle}}^{\nu\mu\rho}}.\label{eq:TotalBelinfanteRosenfeldEnergyMomentumTensor}
\end{align}
 In the free-field limit\textemdash meaning in the absence of the
particle\textemdash this energy-momentum tensor reduces to the standard
gauge-invariant Maxwell energy-momentum tensor, as expected: 
\begin{equation}
T^{\mu\nu}=\frac{1}{\mu_{0}}F^{\mu\rho}\tud F{\nu}{\rho}-\eta^{\mu\nu}\frac{1}{4\mu_{0}}F^{\rho\sigma}F_{\rho\sigma}.\label{eq:FreeFieldBelinfanteRosenfeldEnergyMomentumTensor}
\end{equation}

\section*{Acknowledgments}

J.\,A.\,B. has benefited from personal communications with Gary
Feldman, Howard Georgi, Andrew Strominger, Bill Phillips, David Griffiths,
David Kagan, David Morin, Logan McCarty, Monica Pate, Alex Lupsasca,
and Sebastiano Covone.

\bibliographystyle{1_home_jacob_Documents_Work_My_Papers_Magnetic_Forces_Can_Do_Work__2020__custom-abbrvunsrturl}
\bibliography{0_home_jacob_Documents_Work_My_Papers_Bibliography_Global-Bibliography}

\end{document}